\documentstyle[12pt]{article}
\def\be{\begin{equation}}
\def\ee{\end{equation}}
\def\a{\alpha}

\def\e{\varepsilon}

\def\la{\lambda}

\def\ph{\varphi}

\def\d{\delta}

\begin{document}
\renewcommand{\theequation}{\thesection.\arabic{equation}} 
 \title{Perihelion precession for modified Newtonian gravity} 
 \author{Hans-J\"urgen  Schmidt} 
\date{June 14, 2008}
\maketitle
\centerline{Institut f\"ur Mathematik, Universit\"at Potsdam}  
\centerline{Am Neuen Palais 10, D-14469 Potsdam, Germany}  
\centerline{e-mail: hjschmi@rz.uni-potsdam.de} 
\begin{abstract}
We calculate the perihelion precession $\delta $ for nearly circular orbits 
in a central potential $V(r)$. Differently from other approaches to this 
problem, we do not assume that the potential is close to the Newtonian 
one. The main idea in the deduction is to apply the underlying symmetries 
of the system to show that $\delta $ must be a function of  $r \cdot V''(r)/V'(r)$,  and 
to use the transformation behaviour of $\delta $ in a rotating system of reference. 
This is equivalent to say, that the effective potential can be written in 
a one-parameter set of possibilities as sum of centrifugal potential and 
potential of the central force. We get the following universal formula
 valid for $V'(r) > 0$
$$
\delta (r) = 2 \pi \cdot 
\left[ \frac{1}{\sqrt{3+ r \cdot V''(r)/V'(r)}} -1 \right] \, .
$$
It has to be read as follows: a circular orbit at this value $r$ exists and 
is stable if and only if this $\delta $ is a well-defined real; and if this is the
case, then the angular difference from one perihelion to the next one 
 for nearly circular orbits at this $r$  is exactly $2 \pi + \delta (r)$. 
Then we  apply this result to  examples of recent interest like modified 
Newtonian gravity and linearized fourth-order gravity.  \par
In the second part of the paper, we generalize this universal 
formula to static spherically symmetric space-times
$$
ds^2 = - e^{2\lambda(r)}dt^2  + e^{2 \mu(r)} dr^2 + r^2 d\Omega^2\, , 
$$
 for orbits near $r$ it reads
$$
\d = 2 \pi \cdot \left[\frac{ e^{ \mu(r)} }{
\sqrt{ 3 - 2r \cdot \lambda'(r)  + r \cdot \lambda''(r) / \lambda'(r)   }}
 - 1 \right] 
$$
and can be applied to a large class of theories. \par 
For  the  Schwarzschild black hole with mass parameter $m>0$ it leads to 
$$
\delta = 2  \pi \cdot \left[ \frac{1}{\sqrt{1 -  \frac{6m}{r}}}  - 1 \right] \, ,
$$
a surprisingly unknown formula. It represents
 a strict result and is applicable for all values $r > 6m$ and  is in 
good agreement with the fact that stable circular orbits exist for $r > 6m$
only. For $r \gg m$, one can develop in powers of $m$ and gets
the well-known approximation 
$$
\d \approx \frac{6 \pi m}{r} \, . 
$$ 
\end{abstract}

\vspace{5.mm} Keyword(s): Perihelion advance

\section{Introduction}
\setcounter{equation}{0} 

Adkins and McDonnell \cite{s1} ``calculate the precession of Keplerian orbits 
under the influence of arbitrary central-force perturbations.'' For some examples
 including the Yukawa potential  they present  the result as  hypergeometric 
 function. For nearly circular orbits,  they arrive at the formula
  for the perihelion precession $\Delta \theta_p$,  \cite{s1}, eq. (11)
\be\label{ad11}
\Delta \theta_p = - \frac{\pi}{GMmL} \frac{d^2V}{du^2} 
\vert_{u=1/L}
\ee
where $G$ is the gravitational constant, $M$ the mass of the central body,
$m$ the mass of the moving  test body, $L$ the radius of the orbit, 
and $u=1/r$ the inverted radial coordinate. The potential $V$ is the perturbation 
of the Newtonian potential, so the total potential is then given by
$V(r) - GMm/r$.   They mention that this formula eq.  (\ref{ad11})
is ``equivalent to a formula for the nearly-circular precession that has been 
used by  Dvali,  Gruzinov and  Zaldarriaga \cite{s38}.'' \par 
In  the fourth section of  \cite{s1}, the Yukawa potential
 is applied
in the form   \cite{s1}, eq. (31)
\be\label{ad31}
V(r) = \frac{\a}{r} \exp(-r/\la)    \qquad \qquad  \la > 0\, .
\ee
Using the parameter $\kappa = L/\la$ they arrive at \cite{s1}, eq. (33)
\be\label{ad33}
\Delta \theta_p (\kappa, 0)= - \frac{\pi \a}{GMm} \kappa^2 \exp(-\kappa) \, . 
\ee
In  the fifth section of  \cite{s1}, the authors apply the fact, that within this 
approach, the famous general relativistic perihelion advance can be 
reproduced by using the first post-Newtonian correction 
\be\label{art}
V(r) = - \frac{G^2 M^2 m L}{c^2 r^3}
\ee
where $c$ is the light velocity. They arrive at \cite{s1}, eq. (42)
\be\label{ad42}
\Delta \theta_p ({\rm{GR}})=  \frac{6 \pi GM}{c^2 L}
\ee
and they also present  limits for the value of the cosmological constant 
 by comparing theoretical and measured  values of the Mercury 
perihelion advance.  \par 
The authors of \cite{s38} investigate those kinds of theories
which possess a linearized form of the field equation of the type
 $$\left(\Box + f(\Box) \right) g_{ij} = T_{ij}$$
and calculate e.g. the anomalous perihelion precession for this kind of 
theories by perturbations around the Newtonian potential. 
\par

In  \cite{s33}, the perturbations in the cosmological context are calculated
for several scalar-tensor theories of gravitation, and for the different 
conformal transformations  the distinction between the Einstein and the Jordan 
frames have been made. They applied the results also  to calculate an 
effective gravitational constant for measurements within the Solar system.
In \cite{s34} and \cite{s34a}, the authors carefully calculate the 
possible measurable effects of tensor-multi-scalar theories of gravitation,
including the secular rate of perihelion advance. 
\par
Davies \cite{s37} deduces the perihelion precession due to 
a perturbing central force on an elliptic Keplerian orbit
 via a perturbation with the Runge--Lenz vector. He mentions
that one can mimic the influence of the outer planets
to the perihelion shifts of the inner ones by replacing each outer
planet by a ring of same total mass, so that the effective potential 
can remain rotationally symmetric.  

\section{Perihelion precession}
\setcounter{equation}{0}
 
A test mass shall move along a periodic orbit in a central potential $V(r)$.  
 We look for the perihelion precession of this orbit. \par
    Without loss of  generality the test mass has unit mass, and the motion takes   
 completely place  in the equatorial plane. We parametrize this plane by 
  $(r, \, \ph)$, denote the  time by
$t$ and use  the dot  to abbreviate for  $d/dt$. Then the Lagrangian reads
 \be\label{A1} 
 L =  \frac{1}{2} \left(  \dot r^2 + r^2 \dot \ph^2  \right) - V(r)\, .   
\ee  
We assume that $V(r)$ is a $C^2$-function at all values $r$ which belong to  
the orbit.  For the orbit $(r(t), \, \ph(t))$ we define perihel and aphel via  
 \be\label{A2} 
 r_1 = \min_{t \in R} \, r(t) \qquad {\rm and} \qquad  r_2 = \max_{t \in R} \, r(t) 
  \ee 
 resp., where $0 < r_1 \le r_2 < \infty$ is assumed. \par  
Let $\ph_0$ be the change of $\ph(t)$ during the change from $r(t)$
 from one perihel to the next aphel. Due to time-reversal invariance, 
the same $\ph_0$ is also the change of $\ph(t)$ from this aphel to the next 
perihel. For $2 \ph_0 = 2\pi$, the orbit is exactly closed after one revolution. 
So it is adequate to define the perihelion precession $ \delta $ by
 \be\label{A3} 
\delta  = 2 (\ph_0 - \pi)\, . 
\ee
For purely radial oscillations, our definition implies $\ph_0 = 0$, i.e.
 $\delta  = - 2 \pi$. If  purely radial motion is  excluded from the  consideration, then  
all values of $\delta $ with $\delta  > - 2 \pi$ may appear as perihelion precession. For 
$\delta  > 0$ we call it perihelion advance. \par
What happens with $\delta $ when the orbit is continuously deformed ? 
Example: Let 
\be\label{A4} 
r(\ph) = 4 + \e  \cos (\ph) + \cos (2\ph) 
\ee
with some parameter $\e <1$. For all values $\e>0$ we get $\delta =0$,
 but at $\e=0$ we get $\delta = -\pi$. This example shows that $\delta $ does 
 not always continuously depend on the orbits. \par
However, in the typical cases,  $\delta $ is a continuous function and for a given fixed 
$V(r)$, we have $\delta (r_1, \, r_2)$, i.e., the prescription of  perihel and aphel 
uniquely  determine  the perihelion precession. We now define 
 \be\label{A5} 
\delta (r_0) = \lim_{r_1, r_2 \to r_0} \delta (r_1, \, r_2)\, .
\ee
The expression $\delta (r_0, \, r_0)$ is formally the perihelion precession
 of an exact circular orbit which does not make any sense. So, what 
is the interpretation of the  limit in eq. (\ref{A5})? It is just the 
perihelion precession of nearly circular orbits which should be well-defined
for those cases where the related exact circular orbit is a stable one. 
\par
It is the purpose of the present paper to deduce several formulas for the
calculation of $\delta $ and to apply them to modified Newtonian gravity. 

\section{Nearly circular orbits}
\setcounter{equation}{0} 

How can we calculate the perihelion precession $\delta (r_0)$ for the 
 nearly circular orbit at $r=r_0$? As we have a second order equation of motion, 
it should be a function of $r_0$, $V(r_0)$, $V'(r_0)$ and $V''(r_0)$ only,
where the dash denotes $d/dr$.  An exact circular orbit at this $r$-value is possible
 if and only if  the repelling centrifugal force is compensated by an attractive central 
force, i.e., if  $V'(r_0) > 0$. \footnote{This sentence is included to fix the sign 
convention and to make clear, that   $V'(r_0) $ may be written in the 
denominator.} Now we start to simplify the problem:  
Adding a constant to the potential  does not alter the orbits, so no 
dependence on $V(r_0)$ should appear. Similarly we argue as follows: 
if we multiply the function $V$ by a positive constant, then we can compensate 
this by multiplying $t$ also by a suitable positive constant without changing 
the orbits, therefore the dependence
on the potential can only be in the form of an expression like
$$
\frac{V''(r_0)}{V'(r_0)} =  [ \ln V'(r_0)]'
$$
which is invariant with respect to a multiplication of $V$ by a positive constant.
Finally, we know that $\delta $ is dimensionless, and here we need the last 
possible argument, $r_0$,  to produce  a dimensionless quantity from it: we define 
 \be\label{B1} 
\hat q = \frac{r_0 \cdot V''(r_0)}{V'(r_0)}\, .
\ee
Example: We assume $ V(r) = - 1/r$, then $\hat q = -2$ according to 
 eq. (\ref{B1}); this potential is the exact Newtonian gravitational field,
where we know that all the bounded orbits are exact ellipses with the
center $r=0$ being located at one of their focal points, so we get 
$\delta  =0$  for this case. This motivates the definition $q = \hat q + 2$, i.e.,
 \be\label{B2} 
 q = \frac{r_0 \cdot V''(r_0)}{V'(r_0)} \, + \, 2  \, .
\ee
Then it holds: $\delta $ must be a function of $ q$. As no other dependencies exist,
 it must be a universal function $\delta  [ q ] $ being valid for all potentials,
and $\delta  [ 0 ] = 0 $ because for the Newtonian theory, $q=0$. \par
 The next step is to find out the exact form of this universal function. 
 A first idea to assume exact linearity in $q$ is not justified, because
then the restriction $\delta  > - 2\pi$ deduced in the previous section would not 
 be realized. \par

To find the exact form of this universal function it
suffices to insert a non-trivial one-parameter set of examples for which the 
solution is known. \par 

To this end  we  discuss how  the perihelion advance changes 
 if $\ph$ is replaced by $\tilde \ph = k \cdot \ph$  with an   arbitrary 
positive parameter $k$, but $r$ remains unchanged. 
We get $\tilde \ph_0 = k \cdot \ph_0$   and with  eq. (\ref{A3}) then 
\be\label{B6} 
\tilde \delta  =   - 2\pi + k \cdot (\delta  + 2 \pi) =
k \cdot \delta  + 2 \pi \cdot  \left( k - 1 \right)\, . 
\ee
The set of possible $\delta $-values is restricted by  $\delta  > - 2\pi$, and by 
eq. (\ref{B6}), also $\tilde \delta  > - 2\pi$. For $k=1$ we get, of course, 
$\tilde \delta  =\delta $.  To find the related $q$-values  we need the equation  
 of motion which is deduced in the next section. 

\section{The  equation  of motion}
\setcounter{equation}{0} 

The Lagrangian   eq. (\ref{A1}) reads
$$
 L =  \frac{1}{2} \left(  \dot r^2 + r^2 \dot \ph^2  \right) - V(r) 
 \, .   
$$  
The angular momentum $M$ is a conserved quantity:  
\be\label{C1} 
M = \frac{\partial L}{\partial \dot \ph} = r^2 \dot \ph \, , \qquad 
{\rm hence} \qquad \dot \ph = \frac{M}{r^2}\, . 
\ee
Radial motion  is already excluded, so $M \ne 0$. Without loss of  generality 
we assume $M > 0$, otherwise we would change  the orientation of 
the $r-\ph-$plane. 
\par
The energy  $E$ is also a conserved quantity:  
 \be\label{C2} 
E =  \frac{1}{2} \left(  \dot r^2 + r^2 \dot \ph^2  \right) + V(r) \, .  
\ee
Inserting  eq. (\ref{C1}) into  eq. (\ref{C2}) we get 
\be\label{C3} 
E =  \frac{  \dot r^2}{2}  + V(r) + \frac{ M^2}{2r^2}\, .  
\ee
We derive  eq. (\ref{C3}) with respect to $t$ and divide by $\dot r$ 
afterwards. Then we get the Newtonian force equation 
 \be\label{C4} 
0 = \ddot r + V'(r) - \frac{ M^2}{r^3}\, ,
\ee
the term with $M^2$ represents the centrifugal force. A circular orbit implies 
$\ddot r =0$, so by  eq. (\ref{C4}),   this is possible at $r=r_0$ for $V'(r_0) > 0$ 
only. \par

To evaluate stability, we define the effective potential as  usual:  
\be\label{C4d} 
V_{\rm eff}(r) = V(r) +  \frac{M^2}{2r^2}
\ee
leading 
to 
\be\label{C4c} 
V_{\rm eff}'(r) = V'(r) - \frac{M^2}{r^3}  
\ee
 and
\be\label{C4c2} 
  V_{\rm eff}''(r) = V''(r) + \frac{3 M^2}{r^4}\, . 
\ee
 A circular orbit at $r=r_0$  requires $V_{\rm eff}(r_0) =E$ due to 
 eqs. (\ref{C3}),   (\ref{C4d}) and  $V_{\rm eff}'(r_0) = 0$ due to eqs.
 (\ref{C4}),  eq. (\ref{C4c}). This implies
$$
M(r_0) = \sqrt{r_0^3 \cdot V'(r_0)}
$$
and
$$
E(r_0) = V(r_0) + \frac{r_0}{2} \cdot V'(r_0)\, .
$$
A simple calculation shows that the following four inequalities are all 
equivalent to  each other:
$$
\frac{dM (r_0)}{dr_0} > 0 \, , \qquad
\frac{dE (r_0)}{dr_0} > 0\, ,  \qquad  V_{\rm eff}''( r_0) > 0
$$
and
\be\label{C4b} 
r_0 \cdot V''(r_0)  + 3 \cdot V'(r_0) > 0\, . 
\ee
A perturbation of the circular orbit can be parametrized by 
slightly changed initial conditions, or equivalently by slightly changed
values of $M$ and $E$. In a first step we restrict to perturbations
which have the same angular momentum $M$ and a slightly changed energy 
 $\tilde E$ instead of $E$. So we have to solve
\be\label{C4k} 
\tilde E =   \frac{  \dot r^2}{2}+  V_{\rm eff}(r)\, .
\ee
To get solutions one needs  $\tilde E > E$. Thus the problem 
 is now equivalent to  a one-dimensional motion in the potential
$V_{\rm eff}$. From eq. (\ref{C4k}) we get
$$
\dot r = \pm \sqrt{2} \cdot \sqrt{\tilde E -   V_{\rm eff}(r)}\, .
$$
Together with $\dot \ph = M/r^2$ we find
$$
\frac{d\ph}{dr} = \frac{\dot \ph}{\dot r} = \pm
\frac{M}{r^2 \cdot \sqrt{2} \cdot \sqrt{\tilde E -   V_{\rm eff}(r)}}\, .
$$
If the 
equation $\tilde E =   V_{\rm eff}(r)$ has two solutions 
$r_1$, $r_2$ near $r_0$ with $r_1 < r_0 < r_2$ we get
$$
\ph_0 = \frac{M}{\sqrt{2}} \cdot \int_{r_1}^{r_2} 
\frac{dr}{r^2  \cdot \sqrt{\tilde E -   V_{\rm eff}(r)}}\, .
$$
In the limit $\tilde E \to E$ we have $r_1, r_2 \to r_0$. We need a positive 
finite value for $\ph_0$ in this limit, and this is possible for
 $V_{\rm eff}''(r_0) > 0$ only, i.e., if  inequality (\ref{C4b}) is valid. 
 If this is fulfilled, then $V_{\rm eff}$ has a regular 
quadratic minimum at $r=r_0$ and the limit value of $\ph_0$
depends on $V_{\rm eff}''(r_0) $ only, not on any higher
 derivatives of $V(r_0)$. This strictly confirms the assumption made above
that the universal formula for $\delta $ does not depend on derivatives of $V$
higher than the second one.\footnote{As it represents a key point in the 
deduction, we give also  the  idea for a   third independent proof 
of  this statement; it is meant as pedagogic remark:
 if one considers the analogous problem of motion in a 
4-dimensional pseudo-Riemannian space-time, then the circular 
orbits are represented by such geodesics, and the nearly circular
orbits are represented by the geodesic deviation equation, which itself
has the components of the curvature tensor as coefficients, i.e., no more
than second derivatives of the potentials appear; and our classical problem 
of motion can be given as an adequate   limit of space-times.} 
\par
In a second step we should also look for perturbations
where $M$ is slightly changed to $\tilde M$. However, such perturbations
can be, due to $dM (r_0)/dr_0 > 0$, rearranged to be perturbations 
at a slightly changed circular orbit with adequately chosen $\tilde r_0 $
instead of $r_0$, so this does  not lead to new conditions. 
\par
Now let $\left( r(t), \, \ph(t) \right)$ be a periodic solution and $k>0$
 a parameter. We define $\tilde r(t) = r(t)$ and $\tilde \ph(t) = k \cdot \ph(t)$.
We look for a potential $\tilde V(r)$ such that 
 $\left( \tilde r(t), \, \tilde  \ph(t)  \right)$ becomes a solution. With    
eq. (\ref{C1}) we get 
 \be\label{C5} 
\tilde M = k \cdot M
\ee
and with  eq. (\ref{C3})
 \be\label{C6} 
\tilde E =  \frac{  \dot r^2}{2}  + \tilde V(r) + \frac{\tilde M^2}{2r^2}\, .  
\ee
An additive constant to the energy can be compensated by adding a constant 
to the potential, so we may assume $\tilde E = E$. A comparison 
of 
 eq. (\ref{C3}) with   eq. (\ref{C5}) and  eq. (\ref{C6})  leads to 
$$
 \tilde V(r) + \frac{k^2 \cdot M^2}{2r^2} = V(r) + \frac{ M^2}{2r^2}
$$
i.e. to 
\be\label{C7} 
 \tilde V(r) = V(r) +  \frac{(1-k^2)\cdot M^2}{2r^2}\, . 
\ee
This means:  here we  used  the transformation behaviour 
of $\delta $ in a rotating system of reference, which  is equivalent to say, 
that the effective potential can be written in a one-parameter set 
of possibilities as sum of centrifugal potential and potential of the central force,
and so the knowledge about $\delta $  for one element of this set 
suffices to calculate it for all other elements of this one-parameter set.
\footnote{This is the key point of the deduction: we give as input only the knowledge 
of $\d$ for the Newtonian potential at $r_0=1$, then by this one-parameter set 
 of  transformations we produce the knowledge about $\d$ for a 
one-parameter class of potentials, and this knowledge suffices to 
identify the universal function $\d[q]$ uniquely.} 
 \par
Example: Let $V(r) = -1/r$, i.e., again the Newtonian potential, we
consider nearly circular orbits at $r_0=1$. Then $V'(r)= 1/r^2$ and with 
 eq. (\ref{C4}) we get using $\ddot r = 0$ just $M=1$. 
Inserting this into  eq. (\ref{C7}) we get
\be\label{C8} 
 \tilde V(r) = - \frac{1}{r} +  \frac{1-k^2}{2r^2}\, . 
\ee
Now we apply  eq. (\ref{B6}). For $k=1$ we have, of course, 
$\delta  =0$, so we get 
\be\label{C9} 
\tilde \delta  = 2\pi \cdot (k-1) \, . 
\ee
With the help of  eq. (\ref{B2}) and   eq. (\ref{C8}) we calculate at
$r_0=1$
$$
\tilde q = \frac{\tilde V''(1)}{\tilde V'(1)} + 2 = \frac{1}{k^2} - 1\, .
$$
All values $\tilde q $ with $\tilde q > - 1 $ can appear this way. We invert this
equation and get with the assumption $k > 0$ the result
\be\label{C10} 
k = \frac{1}{\sqrt{1 + \tilde q}} \, .
\ee
Inserting  eq. (\ref{C10}) into  eq. (\ref{C9})  and
 removing the tilde
we get for arbitrary values $q > -1$
\be\label{C11} 
\delta  [ q ] = 2 \pi \left(  \frac{1}{\sqrt{1+q}} -1 \right)\, .
\ee
Indeed, this $\delta $ covers all values $\delta  > - 2\pi$ and it is defined for all values
 $q > -1$. Applying eq. (\ref{B2})  and the condition  $V'(r_0)> 0$  
already mentioned in section 3 this is equivalent to the conditions
\be\label{C12}
r_0 \cdot V''(r_0)  + 3 \cdot V'(r_0) > 0 \qquad {\rm and} \qquad V'(r_0)> 0\, . 
\ee 
 \par   
Equation (\ref{C11}) together with eq.  (\ref{B2}) defines the universal function 
we had looked for in section 2. It should be emphasized that we did not solve any 
integrals or  differential equations to deduce it, we only applied the 
obvious symmetries of the system and the knowledge about the absence 
of perihelion precession in Newtonian gravity.  \par
It is still unclear what conditions for the potential $V(r)$ have to be met that
a periodic orbit with prescribed values of perihel $r_1$ and aphel $r_2$ 
exists. To this end let us fix a $C^2$-function $V(r)$ and values 
$r_1$, $r_2$ with $0<r_1<r_2$. Both  at  $r_1$ and $r_2$ we have 
$\dot r =0$, so we get from  eq. (\ref{C3}) 
\be\label{C13}
V(r_1) + \frac{M^2}{2r_1^2} = V(r_2) + \frac{M^2}{2r_2^2} \, . 
\ee
So one needs the finite version of the condition $V'(r_1)> 0$: 
\be\label{C14}
\Delta V = V(r_2) - V(r_1) > 0\, . 
\ee
By the way, the purely radial oscillations which are excluded here, 
 appear as $M=0$ in  eq. (\ref{C13}) and require $\Delta V = 0$. \par
Inserting  eq. (\ref{C14})  into eq. (\ref{C13}) and solving for $M$ we get
\be\label{C15}
M = r_1 r_2 \sqrt{  \frac{2 \Delta V}{r_2^2 - r_1^2}  } \, .
\ee
Inserting eq. (\ref{C15}) into eq. (\ref{C3}) at $r=r_1$  we get
\be\label{C16}
E =   \frac{r_2^2 V(r_2) -  r_1^2 V(r_1) }{r_2^2 - r_1^2}\, .
\ee
That the motion  between $r_1$ and $r_2$ is always possible requires
 $\dot r \ne 0$ in this whole interval,
\footnote{A priori, an inflexion point
might be  possible, too, i.e., for instance $\dot r > 0$ on an
interval but a single point where $\dot r = 0$. However, such a
behavior is already excluded by our assumption that we 
consider only periodic orbits; a little bit more sophisticated one can argue:
such a behaviour  would a priori allow simultaneously two solutions of the
field equations, showing that the Cauchy problem would be ill-defined
at this point.}
 so we have to fulfil, see eq. (\ref{C3})
\be\label{C17}
E > V(r) + \frac{M^2}{2r^2} \qquad {\rm for} \qquad r_1<r<r_2 \, .
\ee
Inserting eqs. (\ref{C15}) and (\ref{C16}) into this inequality we get
\be\label{C18}
V(r) <    \frac{r_2^2  \left(r^2 - r_1^2 \right) V(r_2) +  r_1^2 
  \left(r_2^2 - r^2 \right) V(r_1) }{r^2  \cdot \left(r_2^2 - r_1^2 \right)}
\ee
representing  the finite version of the first of the  conditions  (\ref{C12}); 
one can prove this statement by inserting $r_2 = r_1 + \e$ into 
eq. (\ref{C18}) and applying the limit $\e \to 0$ afterwards. \par
Finally it should be mentioned that in some limiting cases, also
equality instead of  a  $<$-relation could lead to some solutions; however,
in those cases either $V(r)$ fails to be a $C^2$-function or 
 the test mass would need infinite time to reach the limit, however 
 this fails to  represent a periodic motion: 
and both is excluded from our considerations.

\section{Nearly circular orbits -- second round}
\setcounter{equation}{0} 

Now we are ready to formulate the result: Let $V(r)$ be a $C^2$-function
and let $r_0>0$ be a fixed value of the radial coordinate. Then an
exact circular orbit at this $r$-value is possible if and only if  the repelling 
centrifugal force is compensated by an attractive central force, i.e., if  
$V'(r_0) > 0$, where the dash at $V$ denotes $d/dr$. This orbit represents a 
stable one in the sense that small perturbations of the initial conditions always 
lead to periodic oscillations around $r=r_0$, if  and only if 
$r_0 \cdot V''(r_0)  + 3 \cdot V'(r_0) > 0$. If both inequalities are fulfilled, 
then the perihelion precession $\delta (r_0)$ of the nearly circular
orbits at $r =r_0$ is well-defined and can be calculated  by use of 
 eqs. (\ref{B2}) and  (\ref{C11})  to 
\be\label{D1}
\delta (r_0) = 2 \pi \cdot 
\left( \frac{1}{\sqrt{3+ r_0 \cdot V''(r_0)/V'(r_0)}} -1 \right)\, .
\ee
This equation represents an exact result  and is not restricted to potentials
close to the Newtonian one. \par
 If we rewrite eq. (\ref{D1}) in the form
\be\label{D1a}
\delta (r_0) = 2 \pi \cdot \left( \frac{1}{\sqrt{3+ r_0 \cdot [ \ln V'(r_0)]' }} -1 \right)\, .
\ee
then imposing the validity of the  inequalities  (\ref{C12}) is equivalent to
require that eq.   (\ref{D1a})  represents a well-defined real function. 
This fact can be explained as follows: In the deduction  of  eq.   (\ref{D1a})
 we only used symmetry arguments and continuous deformations of the 
orbits, so in the connected component of the Newtonian potential 
$V(r) = - 1/r$  with $V'(r) > 0$ all regularity conditions will be met; as 
$V'(r_0) =0$ represents a singular point for eq.   (\ref{D1}), it is
also clear from that version of the equation why only $V'(r) > 0$ is 
allowed. 
\par
To ease comparison with the literature it proves useful to work in the 
inverted radial coordinate $u= 1/r$. We define $W(u) = V(r)$ and a dash
at $W$ shall denote $d/du$. Then it holds
\be\label{B3} 
W'(u) = - r^2 \cdot V'(r)\, , \qquad  W''(u) = 2r^3 \cdot V'(r)
 + r^4 \cdot V''(r)\, .
\ee
Because this inversion is a dual  transformation we can  exchange $u$ with 
$r$ and  simultaneously $V$ with $W$ in  eq. (\ref{B3})  and get
\be\label{B4} 
V'(r) = - u^2 \cdot W'(u)\, , \qquad  V''(r) = 2u^3 \cdot W'(u)
 + u^4 \cdot W''(u)\, .
\ee
Combining  eq. (\ref{B2}) with eq. (\ref{B4})  we get  at $u_0 = 1/r_0$
\be\label{B5} 
q=  - \frac{u_0 \cdot W''(u_0)}{W'(u_0)} \, . 
\ee
Then we get with  eq. (\ref{D1})
\be\label{D2}
\delta (1/u_0) = 2 \pi \cdot 
 \left( \frac{1}{\sqrt{1-  u_0 \cdot W''(u_0)/W'(u_0)}} -1 \right)\, .
\ee
In this coordinate the inequalities   (\ref{C12}) read, see   eq. (\ref{B3})
\be\label{D3}
u_0 \cdot W''(u_0) - W'(u_0) > 0  \qquad {\rm and} \qquad W'(u_0)< 0\, . 
\ee
Analogously to eq. (\ref{D1a}) we can now combine eq. (\ref{D2})
with inequalities  (\ref{D3}) to get 
 \be\label{D2a}
\delta (1/u_0) = 2 \pi \cdot 
 \left( \frac{1}{\sqrt{1-  u_0 \cdot \left[ \ln \left(
- W'(u_0)\right) \right]'}} -1 \right)\, .
\ee
If we have the additional condition that 
 $ \vert u_0 \cdot W''(u_0) \vert \ll \vert W'(u_0) \vert $, 
then we get from   eq. (\ref{D2}) the following approximation for $\delta $: 
\be\label{D4}
\delta (1/u_0) = - \pi \cdot 
    u_0 \cdot W''(u_0)/ \vert W'(u_0) \vert \, .
\ee
\par The Newtonian potential for a central mass $m>0$ reads 
$V(r) = - mG/r$, where $G$ is the gravitational constant. This leads
to $W(u) = -m G u$, hence $W'(u) = - mG$ and $W''(u)=0$. 
Therefore, eq. (\ref{D4}) is especially useful if the central
potential under consideration is a small perturbation of 
the Newtonian potential. 

\section{Comparing with Adkins and McDonnell}
\setcounter{equation}{0} 

In \cite{s1}, see also the references cited there, the orbital precession 
due to central-force perturbations has been calculated in details, and 
applications are given;  
especially, their eq. (11)  
(i.e. our eq. 
(\ref{ad11}) above) 
$$
\Delta \theta_p = - \frac{\pi}{GMmL} \frac{d^2V}{du^2} 
\vert_{u=1/L}
$$
 is, after adequate 
transformation  of  the notation, almost  identical to our eq.  (\ref{D4}). \par
 Let us check this statement in more details. To this end we now transform 
our formulas to the notation used in \cite{s1}. This first means: from now
on we  denote the central mass by $M$ and the mass of the test body 
by $m$, and we reintroduce gravitational constant $G$ and light 
velocity $c$ into the formulas, even in those cases, where units have been
chosen with $G=c=1$. Our $W(u)$ in  eq. (\ref{D4}) has then to be replaced 
by $V(u) - GMmu$, where now this $V(u) $ is the small perturbation
 of the Newtonian potential according to  \cite{s1}. This leads to 
$W'(u) = V'(u) - GMm$ and $ W''(u) =V''(u)$. So, for the second derivative,
 there is no difference. For the first derivative, we have within the used
approximation, that  $V'(u)$ can be neglected in comparison with $GMm$. 
 So far, the r.h.s. of    eq. (\ref{D4}) reads
$$
- \frac{\pi u_0 V''(u_0)}{GMm}
$$
and evaluating this at $u_0 = 1/L$ exactly leads to the r.h.s. of    
eq. (\ref{ad11}). \par 
Example:  
Let $V$ be according to eq. (\ref{ad31}), i.e.
$$
V(r) = \frac{\a}{r} \exp(-r/\la)    \qquad \qquad  \la > 0\, .
$$
Then we have 
\be\label{F1}
W(u) = - GMmu \cdot [1 - \beta \exp(-1/(\la u)) ]
\ee
with $\alpha = GMm \beta$. The perihelion shift according to    \cite{s1} 
is with $\kappa = L/\la$, see eq. (\ref{ad33}): 
\be\label{F2}
\Delta \theta_p = - \pi \beta  \kappa^2 \exp(-\kappa) 
\ee
an expression which is, as a consequence of the approximation used, 
completely linear in the parameter $\beta$. In fact, the result is valid
only in regions, where the perturbation is sufficiently small. \par
Now we apply our formula  eq. (\ref{D2}) to the same problem
eq. (\ref{F1}). The factor $GMm$ will cancel out anyhow in  
  eq. (\ref{D2}), so we may put  $GMm = 1$, i.e. $\a = \beta$, already
now. We get 
 \be\label{F3}
W(u) = -  u +  \beta u \cdot  \exp(-1/(\la u)) \, , \quad 
W'(u) = - 1 + \left( 1 + \frac{1}{\la u} \right)   \cdot  \exp(-1/(\la u)) \, . 
\ee
The  more conventional form of this potential $W$ appears 
when one writes
it in dependence on $r=1/u$ to get
 \be\label{F3a}
W = - \frac{1}{r} + \frac{ \beta}{r} \cdot  \exp(-r/\la ) \, . 
\ee
In the present case, the second inequality (\ref{D3}) evaluated at $u_0 = 1/L$
 and using $\kappa = L/\la > 0$ reads
\be\label{F4}
\beta \cdot  (1 + \kappa) < \exp(\kappa)
\ee
and gives a restriction for $\beta > 1$ only. In details: let us fix any 
$\kappa_0 >0$, then we define 
$$
\beta =\exp(\kappa_0) /    (1 + \kappa_0)
$$
and then (\ref{F4}) is fulfilled for $\kappa > \kappa_0$ only. \par
 In other words: For every $\beta \le 1$, all positive radius
 values $L$ appear for a circular orbit. For each $\beta > 1$,
there is a positive $r_0$ such that a circular orbit 
exists for $L > r_0 $ only. 
\footnote{Here, this example was chosen mainly  to present how to apply 
our formula. However, if one wants to interpret the physics behind,
one should note that positive values of  $\beta$  would
correspond to ghost degrees of freedom if they have an even
helicity. But if the extra Yukawa force is mediated by a vector field
(like the W and Z bosons), then even positive values of  $\beta$ are
allowed.}
\par 
For the  second derivative we get from eq. (\ref{F3}):
 \be\label{F5}
W ''(u) = \frac{ \beta }{  \la^2u^3    }  \exp(-1/(\la u)) \, .
\ee
Inserting eqs. (\ref{F3}) and  (\ref{F5}) into eq. (\ref{D2}) we get 
 \be\label{F6}
\delta(L) = 2 \pi \cdot \left(  \frac{1}{\sqrt{1 - 
\frac{\kappa^2}{1 + \kappa - \exp(\kappa)/\beta}
}} -1 \right) \, . 
\ee
Let us examine eq. (\ref{F6}): For very small values $\vert  \beta \vert \ll 1$
it is continuous and eq. (\ref{F2}) is a good approximation to it in 
 correspondence with the fact, that here the perturbation to the Newtonian
potential  is small and therefore,  eq. (\ref{F2}) is applicable. 
\par
Things are quite different for other cases: From  eq. (\ref{F3a}) one can see that 
only for $\beta =1$, the potential remains bounded as $r \to 0$. For this case and
small values of $\kappa \ll 1$ eq.   (\ref{F2})  gives $\d = - \pi\kappa^2$, whereas 
the exact formula  eq. (\ref{F6}) has in the same case $\d =  - 2\pi(1 - 1/\sqrt{3})$,
 a totally different behaviour. 
\par 
In section 3 of   \cite{s1} it is mentioned that   the development of  
perihelion precession in powers of the eccentricity $e$ contains only
even powers of $e$; this means that for sufficiently small values of $e$,
where linearization in $e$ is justified, our formulas for nearly 
circular orbits are applicable, too.

\section{Application to fourth-order gravity}
\setcounter{equation}{0}

Further examples are as follows: 
In  \cite{s2} I wrote without explicit proof, see also \cite{s3}, page 235-236:
``Next,  let us study the perihelion advance for distorted circle-like orbits. 
Besides the general relativistic perihelion advance, which vanishes 
in the Newtonian limit,  we have an additional  one of the following 
behaviour: for $r \to   0$  and $ r \to \infty $  it vanishes and for  $r \approx 1/m_0$  
and $r \approx 1/m_2$  it has local maxima,  i.e., resonances.''  \par
 This refers to linearized fourth-order gravity, see  Stelle  \cite{stelle78} 
for details, where the gravitational potential for a point mass $m$ reads 
\be\label{12.15} 
V(r) = - m r^{-1} \bigl( 1 + \exp (-m_0 r)/3 - 4 \exp (-m_2 r)/3 \bigr) \, .
\ee
The perihelion precession of this and similar theories can be calculated
by inserting this potential $V(r)$ into the equation (\ref{D1}). Of course,
 in the region of large values $r$, the known approximations like
(\ref{D4}) would serve also, but our equation (\ref{D1}) will give
 the correct result also for those $r$-values,  where $V(r)$  is far from 
being close to the Newtonian potential. \par 
Here in  eq. (\ref{12.15}), $m_0$ is the mass of  
the massive spin 0-graviton stemming from 
the $R^2$-term in the Lagrangian, and $m_2$ is the mass of  the massive 
spin 2-graviton \footnote{This massive spin-2 excitation  is a ghost, i.e., 
carries negative kinetic energy and thereby spoils the stability of the
model. Therefore, this contribution  has  not a direct  phenomenological
interpretation.}
 stemming from the term $C_{ijkl} C^{ijkl}$ in the Lagrangian.
Both $m_0$ and $m_2$ are assumed to be positive to exclude the 
appearance of tachyons, but $m_0 \to \infty$ and $m_2 \to \infty$
 represent sensible limits. \par
In the case $m_0 = m_2 > 0$,  eq.  (\ref{12.15}) exactly leads to the 
case $\beta = 1$ discussed in the previous section, the other cases are similar. 

\section{Discussion -- first part}
\setcounter{equation}{0} 

 At that time paper \cite{s2}
 was first published, in 1986, there this was a purely theoretical
question. However, recently there is a development to take such quadratic 
gravity theories quite seriously in the sense that their predictions can 
be confronted with observations, see e.g. \cite{s13}, \cite{s16},
 \cite{s16a}, \cite{s21}, \cite{s17} and the references
 cited there. Also the cosmological solutions of this kind of theories have
been analyzed in more new details recently, see  e.g. \cite{s14}, \cite{s15},
\cite{s18}, \cite{s19},   \cite{s20}, \cite{s22}, \cite{s23},  \cite{s24},
 \cite{s25}, \cite{s26},   \cite{s27}, \cite{s28} and \cite{s29}. \par 
Further, it should be mentioned, that for very distorted orbits, 
\cite{s30} and \cite{s31} give exact results for the perihelion precession 
for a perturbed Newtonian potential.  \par
Now, let us  look  for which  potentials $V(r)$  the parameter $\delta (r_0)$
 takes  values  according to equation (\ref{D1})
 which do not depend on $r_0$. 
After some calculation we get:  up to the  inessential transformations
mentioned above,  there is a parameter $c>0$ such that 
\be\label{t25}
\delta  = 2\pi (c^{-1/2} -1)
\ee
and
\be\label{t26}
V(r) = \ln r  \quad {\rm for} \quad  c=2\, , \qquad
 V(r) =  \frac{1}{c - 2} \cdot r^{c-2}\quad  {\rm else}.
\ee
As expected, one just gets the self-similar functions as solutions to this problem. 
The case $c=1$, i.e., $\delta  =0$  just recovers the Newtonian case.  
Here the bounded orbits are all exact ellipses, the center of
symmetry of the potential,  $r=0$,  being at one of  their focal points. \par
 For $c=4$, eq. (\ref{t26}) leads to the harmonic oscillator  $V =  r^2/2$;
 here the bounded orbits are  also ellipses, but now, the center of symmetry 
of the potential coincides with the center of the ellipses, therefore, the next
perihelion is already after one half rotation, i.e. $\ph_0 = \pi$ and $\delta  = - \pi$
 in accordance with eq.  (\ref{t25}) for this case. This result once more 
confirms that our result is a strict one also far from the Newtonian potential. 

To prepare for the next part, we now 
 apply units such that light velocity $c=1$; then it holds: the 
velocity of the test particle in the exact circular orbit at $r=r_0$
 is less than light velocity if 
\be\label{D18}
r_0 \cdot V'(r_0) < 1\, .
\ee

\section{Circular and nearly circular geodesics}
\setcounter{equation}{0} 

In this second part of the paper 
 we generalize the results  of the first part   to static spherically symmetric 
space-times, see also     \cite{s4}, 
     \cite{s5},    \cite{s6} and  \cite{s61}   for other papers on similar topics. \par
Let us now generalize the resulting  eq. (\ref{D1}) to the analogous situation  in 
a 4-dimensional  static spherically symmetric space-time. 
\footnote{A deduction fully analogous to that
one from the first part seems not to be easily done. The second variant, 
namely to apply the geodesic deviation equation, also leads to 
unnecessary complicated expressions. As third idea one could try  to apply 
general exact solutions of the geodesic equation as found in  the text-book 
literature, e.g.  \cite{s63}, but the elliptic integrals appearing there 
are not easy to handle, therefore we now choose a fourth method,
namely the direct calculation with nearly circular geodesics. }
 We additionally assume that Schwarzschild coordinates are possible,
 so we consider the   metric
\be \label{H1}
ds^2 = - e^{2\la}dt^2  + e^{2 \mu} dr^2 + r^2 d\Omega^2
\ee
where $d\Omega^2$ is the metric of the standard 2-sphere
 and  $\la$ and $\mu$ depend on $r$ only. 
We look for time-like geodesics in this space-time (\ref{H1}).
 After suitable rotation of the coordinate system this geodesic 
remains completely in  the equatorial plane. Due to the chosen symmetry
it holds: Geodesics in the equatorial plane of  the 4-dimensional 
space-time (\ref{H1})  are exactly the geodesics in the 
3-dimensional space-time
\be\label{H2}
ds^2 = - e^{2\la}dt^2  + e^{2 \mu} dr^2 + r^2 d\ph^2 \, .
\ee
The coordinates in (\ref{H2}) are $x^i$ where $i=0$, 1, 2, and the
geodesic shall be parametrized by its natural parameter $\tau$:
\be\label{H3}
x^i (\tau) = \left( t (\tau), \, r (\tau), \, \ph (\tau) \right) \, .
\ee
 With a dot denoting  $d/d\tau$ we get from   (\ref{H2})
\be\label{H4}
-1 =  - e^{2\la}\dot t^2  + e^{2 \mu} \dot r^2 + r^2 \dot \ph^2\, . 
\ee
We may assume $\dot t > 0$. The  components of the geodesic equation 
read:  
\be \label{H5}
0 = \ddot t + 2  \la'  \,  \dot t \,  \dot r \, , 
\ee
\be \label{H6}
0 = e^{2 \mu} \left(  \ddot r +  \mu' \,  \dot r^2 \right)
+  \la'\,  e^{2\la}\,  \dot t^2 - r \dot \ph^2
\ee
and
\be \label{H7}
0 = r \ddot \ph + 2\,  \dot r \dot \ph\, . 
\ee
We define angular momentum $M$ as usual: 
\be \label{H8}
M = r^2 \dot \ph \, .
\ee
Purely radial motion shall not be considered, so we  have $M \ne 0$. Without 
loss of generality we may assume $M>0$, for otherwise  we could reverse 
the orientation of the equatorial plane. 
\par
 Due to  eq.    (\ref{H7}), $M$ is a conserved quantity,  and we 
apply this fact to simplify eqs.    (\ref{H4}) and    (\ref{H6}) to 
\be\label{H9}
-1 =  - e^{2\la}\dot t^2  + e^{2 \mu} \dot r^2 + M^2 / r^2  
\ee
and 
\be \label{H10}
0 = e^{2 \mu} \left(  \ddot r +  \mu' \,  \dot r^2 \right)
+  \la'\,  e^{2\la}\, \dot t^2 -  M^2 / r^3
\ee
resp. Inserting eq. (\ref{H9}) into eq. (\ref{H10}) we can cancel $t$  to get 
\be \label{H11}
0 = e^{2 \mu} \left(  \ddot r + (\la' +  \mu') \,  \dot r^2 \right)
+  \la'\,   \left(1  + \frac{M^2}{r^2} \right)  - \frac{ M^2 }{ r^3}\, . 
\ee
Next we look for a circular orbit at a fixed value $r_0$ of the radial 
coordinate $r(\tau) \equiv r_0 > 0$. With eq.  (\ref{H11}) we get 
\be \label{H12}
r_0 \cdot \la'(r_0) = \frac{z}{z+1}\, , \qquad {\rm where} \qquad
z = \frac{M^2}{r_0^2} > 0\, .
\ee
This means: a circular time-like geodesic 
orbit at $r=r_0$ exists if and only if the inequalities 
\be \label{H13}
 \la'(r_0) > 0 \qquad {\rm and} \qquad 
r_0 \cdot \la'(r_0)  < 1
\ee
are fulfilled. They are fully analogous to the second of the inequalities
  (\ref{C12}) and inequality  (\ref{D18}) resp. \par
 From eq.   (\ref{H12}) we get 
\be \label{H15}
M^2 =  \frac{r_0^3 \cdot \la'(r_0)}{1 - r_0 \cdot \la'(r_0)} \, . 
\ee
For $M$ eq.  (\ref{H15})   the condition  $d  M / dr_0 > 0$ is equivalent to 
\be \label{H15a}
r_0 \cdot \la''(r_0)  + 3 \cdot \la'(r_0) > 2r_0 \left( \la'(r_0)  \right)^2\,  .
\ee
Next, let us define energy $E$ by 
\be\label{H16}
E = e^{2\la(r)} \, \dot t > 0 \, .
\ee
Due to eq.  (\ref{H5}), $E$ is a conserved quantity. We can apply this
equation to remove $t$ from eq.  (\ref{H9}), leading to\footnote{The material
in this section is essentially text-book standard, as can be found e.g. in
  \cite{s63}; but we presented it here in details to maintain a self-consistent 
notation.} 
\be\label{H17}
-1 =  - E^2 \, e^{- 2\la}  + e^{2 \mu} \dot r^2 + M^2 / r^2  \, .
\ee
For circular orbits at $r=r_0$ this leads to 
\be\label{H18}
 E^2 = e^{ 2\la(r_0)} \left(1 + \frac{ M^2 }{ r_0^2} \right)  \, .
\ee
Inserting  eq.  (\ref{H15})  into  eq.  (\ref{H18}) we get
for the energy of the circular orbit at $r=r_0$
\be\label{H19}
E = \frac{e^{ \la(r_0)}}{\sqrt{1 - r_0 \cdot \la'(r_0)}} \, . 
\ee
The condition  $d  E / dr_0 > 0$ is  equivalent to 
the condition   (\ref{H15a}). 

\section{Perihelion precession in space-time}
\setcounter{equation}{0} 
We now prescribe a value $r_0 > 0$ such that (\ref{H13}) and  (\ref{H15a}) 
are fulfilled.  The circular orbit at $r= r_0$ has angular momentum according 
to  (\ref{H15})  and energy $E$  according to  (\ref{H19}).
This circular orbit shall now be perturbed, the perturbed orbit shall have the 
 same  angular momentum $M$ but a different energy $\bar{E} \ne E$. For the 
radial coordinate $r$  in dependence on $\tau$ we make the  following ansatz
\be\label{K1}
r = r_0 + \e \cdot \sin(\a \tau)
\ee
where $\a$ is a positive parameter and $\e$ shall be small such that higher
powers of $\e$ may be neglected. We insert this ansatz   (\ref{K1}) into
eq.   (\ref{H11})  and get the following identity
\be\label{K2}
\a = e^{- \mu(r_0)} \cdot \frac{\sqrt{
r_0 \cdot \la''(r_0)  + 3 \cdot \la'(r_0) - 2r_0 \left( \la'(r_0)  \right)^2
}}{\sqrt{1 - r_0 \cdot \la'(r_0) } \cdot \sqrt{r_0}} \, .
\ee
It is remarkable that just the inequalities deduced before ensure that
$\a$ becomes a well-defined positive real. From eqs.    (\ref{K1})  and  (\ref{K2})
we get:  The time from one perihelion to the next is then $\tau_0$ defined 
by 
\be\label{K3}
\tau_0 = 2 \pi / \a \, . 
\ee
The perihelion shift $\d$ is defined as
\be\label{K4}
\d = \ph(\tau_0) - \ph(0) - 2 \pi\, ,
\ee
it measures how much the change in the angular coordinate $\ph$
differs from $2\pi$ when the orbit changes from one perihelion to the next one. 
Clearly, for $\d =0$, the orbits are exactly closed after one revolution. 
From eqs.     (\ref{H8}), (\ref{K3})  and  (\ref{K4}) we get 
\be\label{K5}
\d = \frac{2\pi}{\a} \cdot \frac{M}{r_0^2} - 2 \pi
\ee
thus leading to the final result  
\be\label{K6}
\d = 2 \pi \cdot \left[\frac{ e^{ \mu(r_0)} }{
\sqrt{ 3 - 2r_0 \cdot \la'(r_0)  + r_0 \cdot \la''(r_0) / \la'(r_0)   }}
 - 1 \right] \, . 
 \ee

\section{Discussion -- second part}
\setcounter{equation}{0} 

The final formula eq.  (\ref{K6}) has a structure quite similar 
to the corresponding formula eq.  (\ref{D1}) from the first part. 
 But a direct change over from one of them to the other 
one is not easily done, so it was really necessary to deduce both 
of them. In eq.  (\ref{K6}) one can observe, that the spatial 
metric component encoded by the function $\mu$ essentially enters 
 the formula but none of their derivatives do enter here. This is 
 in contrast to the temporal metric component encoded by the function 
$\la$ from which only the first and second derivative do enter. 
 \par
As a first test, let us insert the Schwarzschild-de Sitter solution into eq.  (\ref{K6}).
In units where $c=G=1$, we have to insert into  eq.  (\ref{H1})
\be\label{K8}
e^{2\la} =  e^{- 2 \mu} = 1 - \frac{2m}{r} - \frac{\Lambda}{3} \cdot r^2
\ee
where $m>0$ is the mass of the source and $\Lambda \ge 0 $
the cosmological constant. For $\Lambda = 0 $ this leads to 
\be\label{K7}
\d = 2 \pi \cdot \left[ \frac{1}{\sqrt{1 - 6m/r_0}}  - 1 \right] \, . 
 \ee
This is a strict result and is applicable for all values $r_0 > 6m$. 
 Eq.   (\ref{K7}) is surprisingly  unknown up to now. It  is in 
good agreement with the fact that stable circular orbits exist for $r_0 > 6m$
only. For $r_0 \gg m$, one can develop in powers of $m$ and gets
the well-known approximation 
\be\label{K9}
\d \approx \frac{6 \pi m}{r_0}\, .
\ee
Example: for $r_0 = 24m$, the exact equation     (\ref{K7}) leads
to $\d = \pi \cdot (4/\sqrt{3} - 2) \equiv 55,7^0$,  whereas the 
approximation  (\ref{K9}) leads to $\d = \pi/4 \equiv 45^0$. \par 
For $\Lambda > 0$, the condition that a time-like circular orbit exists, is the 
same as for  $\Lambda = 0$, namely $r_0 > 3m$, but 
the formula for perihelion shift becomes a little bit more complicated, 
\be\label{K10}
\d = 2 \pi \cdot \left[ \frac{1}{\sqrt{1 - 6m/r_0}} \cdot  \left(1 + 
\frac{\Lambda r_0^3}{6m} \cdot \left(3 + \frac{9m}{r_0-6m} \right)
\right)  - 1 \right] \, . 
\ee
but in the case $\Lambda << 1/m^2$ and $r_0$ not too large one gets 
the useful approximation
$$
\d \approx \frac{6 \pi m}{r_0} + \frac{\pi  \Lambda r_0^3 }{m}\, .
$$
In a following paper, eq.  (\ref{K6}) shall be applied also to other
spherically symmetric metrics.

\section*{Acknowledgement}

I thank Nathalie Deruelle for posing fruitful questions related
 to perihelion advance in quadratic gravity theories.

\end{document}